\documentclass[prd, superscriptaddress, ctexart, nofootinbib, twocolumn, showpacs]{revtex4}
\usepackage{graphicx}
\usepackage{epsfig}
\usepackage{bm}
\usepackage{verbatim}
\usepackage{amsfonts}
\usepackage[latin1]{inputenc}
\usepackage{graphicx}
\usepackage{amssymb}
\usepackage{color}
\usepackage{float}
\usepackage{amsmath}
\usepackage{amsfonts}
\usepackage{bm}
\usepackage{tikz}

\usetikzlibrary{decorations.pathmorphing}
\newcommand{\beq}{\begin{equation}}
\newcommand{\eeq}{\end{equation}}
\newcommand{\bea}{\begin{eqnarray}}
\newcommand{\eea}{\end{eqnarray}}

\def\be{\begin{equation}}
\def\ee{\end{equation}}
\def\ba{\begin{eqnarray}}
\def\ea{\end{eqnarray}}

\def\be{\begin{equation}}
\def\ee{\end{equation}}
\def\ba{\begin{align}}
\def\ea{\end{align}}

\begin{document}
\title{Searching for a matter bounce cosmology with low redshift observations}
\author{Yi-Fu Cai}
\email[]{yifucai@ustc.edu.cn}
\affiliation{CAS Key Laboratory for Researches in Galaxies and Cosmology, Department of Astronomy, University of Science and Technology of China, Chinese Academy of Sciences, Hefei, Anhui 230026, China}
	
\author{Francis Duplessis}
\email[]{fdupless@asu.edu}
\affiliation{Department of Physics, Arizona State University, Tempe, Arizona 85287, USA}

\author{Damien A.~Easson}
\email[]{easson@asu.edu}
\affiliation{Department of Physics  \& Beyond Center for Fundamental Concepts in Science,\\
Arizona State University, Tempe, Arizona 85287, USA}

\author{Dong-Gang Wang}
\email[]{wdgang@mail.ustc.edu.cn}
\affiliation{CAS Key Laboratory for Research in Galaxies and Cosmology, Department of Astronomy, University of Science and Technology of China, Chinese Academy of Sciences, Hefei, Anhui 230026, China}

\begin{abstract}
The matter bounce scenario allows for a sizable parameter space where cosmological fluctuations originally exited the Hubble radius when the background energy density was small. In this scenario and its extended versions, the low energy degrees of freedom are likely responsible for the statistical properties of the cosmic microwave background power spectrum at large length scales. An interesting consequence is that these modes might be observable only at relatively late times. Therefore low redshift observations could provide evidence for, or even falsify, various bouncing models. We provide an example where a recently hinted potential deviation from  $\Lambda$-cold-dark-matter cosmology results from a dark matter and dark energy interaction. The same interaction allows matter bounce models to generate a red tilt for the primordial curvature perturbations in corroboration with cosmic microwave background experiments.
\end{abstract}


\maketitle
	
\section{Introduction}

The recently released cosmic microwave background (CMB) data from the Planck Collaboration has constrained the value of the spectral index to be $n_s=0.968\pm0.006$ \cite{Ade:2015xua}, verifying at high precision a nearly scale-invariant power spectrum of primordial curvature perturbation with a slightly red tilt. These properties are naturally achieved in inflationary cosmology where a nearly scale-invariant power spectrum is associated with an almost constant Hubble scale during inflation as described by cosmological perturbation theory \cite{Mukhanov:1990me}. Despite the many successes of the inflationary universe paradigm, recent precision observations
are beginning to statistically disfavor many of the simplest (polynomial field potential) models (see, e.g. Planck~\cite{Ade:2015xua}, and the BICEP2/Keck Array ~\cite{Array:2015xqh}). In light of the latest observations
it is worthwhile to continue
to search for potential alternative theories of the early Universe. The study and interpretation of primordial cosmological perturbation theory may also be performed in alternative early Universe paradigms, such as bouncing cosmology \cite{Cai:2014bea, Novello:2008ra, Battefeld:2014uga}, ekpyrotic cosmology~\cite{Khoury:2001wf, Lehners:2008vx}, the pre-big bang model \cite{Gasperini:1992em, Veneziano:1991ek}, and string gas cosmology \cite{Brandenberger:1988aj, Battefeld:2005av, Brandenberger:2015kga}. Among some of these scenarios, it was pointed out in Refs. \cite{Wands:1998yp, Finelli:2001sr} that
a massless scalar field will acquire a scale-invariant power spectrum when its vacuum fluctuations are allowed to exit the Hubble radius during a matter-like (background equation-of-state parameter $w=0$) contracting phase.

The aforementioned scenario, known as {\it matter bounce} cosmology, has been extensively studied in the literature. A challenge matter bounce cosmology has to address is whether the scale-invariant primordial power spectrum can survive the bouncing phase. This issue has been studied in several models, for example, the quintom bounce~\cite{Cai:2007qw, Cai:2007zv}, the Lee-Wick bounce \cite{Cai:2008qw}, the Horava-Lifshitz gravity bounce \cite{Brandenberger:2009yt, Cai:2009in, Gao:2009wn}, the $f(T)$ teleparallel bounce \cite{Cai:2011tc, deHaro:2012zt, Cai:2015emx}, the ghost condensate bounce \cite{Lin:2010pf}, the Galileon bounce \cite{Qiu:2011cy, Easson:2011zy}, the matter-ekpyrotic bounce \cite{Cai:2012va, Cai:2013kja, Cai:2014zga}, the fermionic bounce \cite{Alexander:2014eva, Alexander:2014uaa}, etc.~(see, e.g. Refs. \cite{Brandenberger:2010dk, Brandenberger:2012zb} for recent reviews). It was found in general that on length scales larger than the time scale of the nonsingular bounce phase, both the amplitude and the shape of the power spectrum remain unchanged through the bounce \cite{Quintin:2015rta,Battarra:2014tga}.

Successful models of a nonsingular bounce must satisfy various observational constraints from CMB measurements. In particular, such models must yield a red tilted spectral index while satisfying both the non-Gaussianity constraints \cite{Cai:2009fn}
and the upper bound on the tensor-to-scalar ratio. Compared to the inflationary paradigm, the simplest matter bounce models do fall short of achieving all these requirements; see Refs. \cite{Cai:2014bea,Battefeld:2014uga} for comprehensive reviews. The mentioned shortcomings resulted in various extensions of the original matter bounce paradigm. For instance, one may realize a small deviation from the exact matter contracting phase to obtain the red tilt. This was achieved in the quasi-matter bounce cosmology \cite{deHaro:2015wda, Elizalde:2014uba} where a very specific and tuned form of the scalar field potential was introduced, making the models potentially unnatural. Recently, it was proposed that a $\Lambda$-cold-dark-matter ($\Lambda$CDM) bounce scenario could also solve the tilt problem by simply considering a cosmological constant with an almost pressureless dark matter fluid during the contracting phase \cite{Cai:2014jla}.
However this natural sounding scenario produces too much running of the spectral index, and we will discuss this in Sec. \ref{sec_redtilt}. Finally, obtaining a small tensor-to-scalar ratio while staying in the regime of low non-Gaussianities has been more challenging \cite{Quintin:2015rta}. In the conclusion we will comment on a mechanism that can potentially achieve this and also nicely align with the philosophy of this paper.

The goal of this paper is to determine whether viable models of the matter bounce can be constructed using low energy degrees of freedom (DOF) that can be studied through low redshift observations. This is motivated by the result of Sec. \ref{EscaleCont} which argues that, for a large percentage of the possible matter bounce models, it is the low energy physics which gives the description of the Universe when the CMB modes first exit the horizon. The $\Lambda$CDM bounce scenario would have been an example of this idea but does not agree with the CMB data. Dark energy and dark matter are the two dominant components governing the evolution of the Universe today and in the $\Lambda$CDM model, those two sectors are decoupled and probed indirectly through their gravitational effects. However there exists the possibility of a small, but nonzero, interaction between the dark sector. This scenario is dubbed the interactive dark energy (DE) model (see Refs. \cite{Amendola:1999er, Comelli:2003cv, Wang:2006qw} for earlier literature and Refs. \cite{Copeland:2006wr, Cai:2009zp, Li:2011sd} for comprehensive reviews on DE dynamics). The interaction term between DE and dark matter (DM) could give rise to new features in the formation of the large scale structure (LSS), and the corresponding constraints were studied extensively in the past (for example, see Refs. \cite{Amendola:2001rc, Boehmer:2008av, He:2008tn, Koyama:2009gd, Guo:2007zk, Xia:2009zzb, Li:2013bya} and references therein). In particular, the recent BOSS experiment of the SDSS Collaboration \cite{Delubac:2014aqe} indicates a slight deviation (at $2\sigma$ C.L.) in the expected $\Lambda$CDM value of the Hubble parameter and the angular distance at an average redshift of $z=2.34$. These observational hints could be modeled by $w$CDM \cite{Cardenas:2014jya, Sahni:2014ooa} and interactive DE \cite{Salvatelli:2014zta, Abdalla:2014cla, Valiviita:2015dfa} models. A statistical analysis of these models using the BOSS data was done in Ref. \cite{Costa:2013sva}.

Here, we consider the influence of the DE and DM interaction during the contracting phase assuming that primordial cosmological perturbations were mainly generated by the vacuum fluctuations of a massless scalar field. In our model, the dark interaction modulates the background evolution to yield a small deviation from the exact matter contraction phase when the Universe was dominated by DM before the bounce. We find it is possible to obtain a red tilt for the primordial isocurvature perturbations which can then be transferred into the adiabatic mode.
In addition, because the dark sector interaction term may survive the bounce and influence late-time dynamics
\footnote{However, as discussed in Ref.~\cite{Alexander:2015pda}, undergoing a bounce at very high energies could result in the variation of the physical constants to random values, and thus the value of the prebounce coupling would not "survive." We will assume that such a situation does not occur.}, 
this physical interpretation of the CMB's red tilted power spectrum would be connected to the essence of the DE and DM dynamics measured by astronomical surveys in the late-time Universe. The interaction would produce observable signatures in the LSS of which we may already have discovered hints \cite{DiValentino:2015bja}. This would provide a testable mechanism that explains the CMB measurement of the spectral index in the context of the matter bounce scenario.
	
The paper is organized as follows. We begin with a brief review of the background and perturbation theory of a matter bounce cosmology in Sec. \ref{skeleton} along with the shortcomings of the original scenario. In Sec. \ref{EscaleCont} we perform the analysis of the energy scales that was present at the time the primordial perturbation modes associated with the CMB window were initially exiting the Hubble radius. From this study, we show that a sizable parameter space allows for a very low energy density at that epoch, and hence any new exotic DOF imparting the statistical properties of the CMB perturbations could in principle be probed with late-time observations where they can become more apparent. In Sec. \ref{sec_redtilt}, we first revisit the possibility of generating a red tilt for the power spectrum of primordial perturbations within the $\Lambda$CDM bounce and describe why it fails. Afterward, we show that a tilt can be achieved after assuming an existence of the interaction term between DE and DM which also causes a small deviation from perfect $\Lambda$CDM expansion at late times. We conclude with a discussion in Sec. \ref{conclusion}.

\section{Essentials of bouncing cosmologies}\label{skeleton}
Here we provide a brief overview of general features found in bouncing cosmologies. For details see the reviews \cite{Brandenberger:2012zb,Battefeld:2014uga}.
Bouncing scenarios postulate that the Universe underwent a contracting phase prior to our current expanding phase. As sketched in Fig. \ref{bouncehistory}, this possibility naturally solves the
horizon problem. It can also address the monopole problem if the bounce occurs at an energy scale below the one needed to generate the massive relics; however, the scenario is impartial to the
flatness problem: while curvature becomes more apparent as a universe expands, the opposite occurs in the contracting phase, and curvature becomes progressively
unobservable. The two effects could cancel out but one might have to impose some degree of flatness by hand.

Bouncing models have roughly three distinct periods. There is an initial regime of contraction in which the background becomes dominated by fluids having progressively larger EoS.
This leads to the bouncing phase followed by
a period of reheating leading to an expanding radiation dominated stage.
Here we are not concerned  with the details of the bounce and reheating phase, and will
focus on the period of contraction. We first discuss the evolution of the background in Sec. \ref{contphaseBack},
and then analyze the statistical properties of cosmological perturbations in Sec. \ref{pertreview}.

\subsection{Contracting phase}\label{contphaseBack}
As the name suggests, the matter bounce scenario assumes that the Universe underwent a matter dominated contraction phase. The length of this period depends on the relative energy density of matter and other components having different equation of state (EoS). If no interactions exists, the Hubble parameter evolves according to
\be\label{HubbleEq}
3M_p^2H^2=\frac{\rho_\Lambda^*}{a^0}+\frac{\rho_m^*}{a^3}+\frac{\rho_r^*}{a^4}+\frac{\rho_a^*}{a^6}+\frac{\rho_\phi^*}{a^{3(w+1)}}+ \cdots \,,
\ee
where $\rho^*$ is the energy density of each component when the scale factor is normalized at the present time $a=1$. For our purposes it suffices to consider the following components in Eq. (\ref{HubbleEq}): a cosmological constant, matter, radiation, anisotropies, and a scalar field $\phi$ with an EoS of $w$ respectively.

As the scale factor decreases during the contracting phase, energy densities with larger EoS tend to progressively drive the Hubble expansion.
A major obstacle in any bouncing universe is the anisotropy problem which leads to a Mixmaster universe \cite{MisnerMU}.
Anisotropies have an energy density scaling as $a^{-6}$ due to their EoS being $w=1$ and therefore dominate over all other forms of energy listed in Eq. (\ref{HubbleEq}). The universe becomes increasingly anisotropic and chaotic, making it impossible to track the background evolution. The common way to bypass this problem is to postulate a super-stiff fluid with $w\gg 1$ making the
scale factors evolve very slowly, thereby preventing the anisotropy from grow appreciably. This is known as an Ekpyrotic phase and this stiff fluid dominated Hubble expansion is the only
relevant epoch in the Ekpyrotic scenario \cite{Khoury:2001wf}.

The condition $H=0$ must hold at the bounce point if the Einstein equations are to remain valid.
This requires a fluid with effectively negative energy density  in order to cancel the large contribution from the other
components discussed above. Such scenarios are called nonsingular bounces as they are able to evade singularity theorems and
the entire evolution of the bounce may be tracked analytically in a consistent effective field theory (EFT) framework \cite{Koehn:2015vvy}. On the other hand, singular bounces invoke unknown quantum
gravity effects that are said to resolve the timelike singularity. If these quantum gravity effects occur for a very short
time, one may treat the bounce phase as a singular surface and then attempt to (ambiguously) apply matching conditions to follow the evolution across the bounce.
Recent development in Ref. \cite{Gielen:2015uaa} introduce methods to track the classical paths of perturbations
in the complex plane, and thereby smoothly go through the bounce, avoiding the cosmological singularity.
See Ref. \cite{Battefeld:2014uga} for detailed discussions and examples of both types of bounces.

\begin{figure}[tbhp]
 	\centering
 	\includegraphics[width=0.35\textwidth]{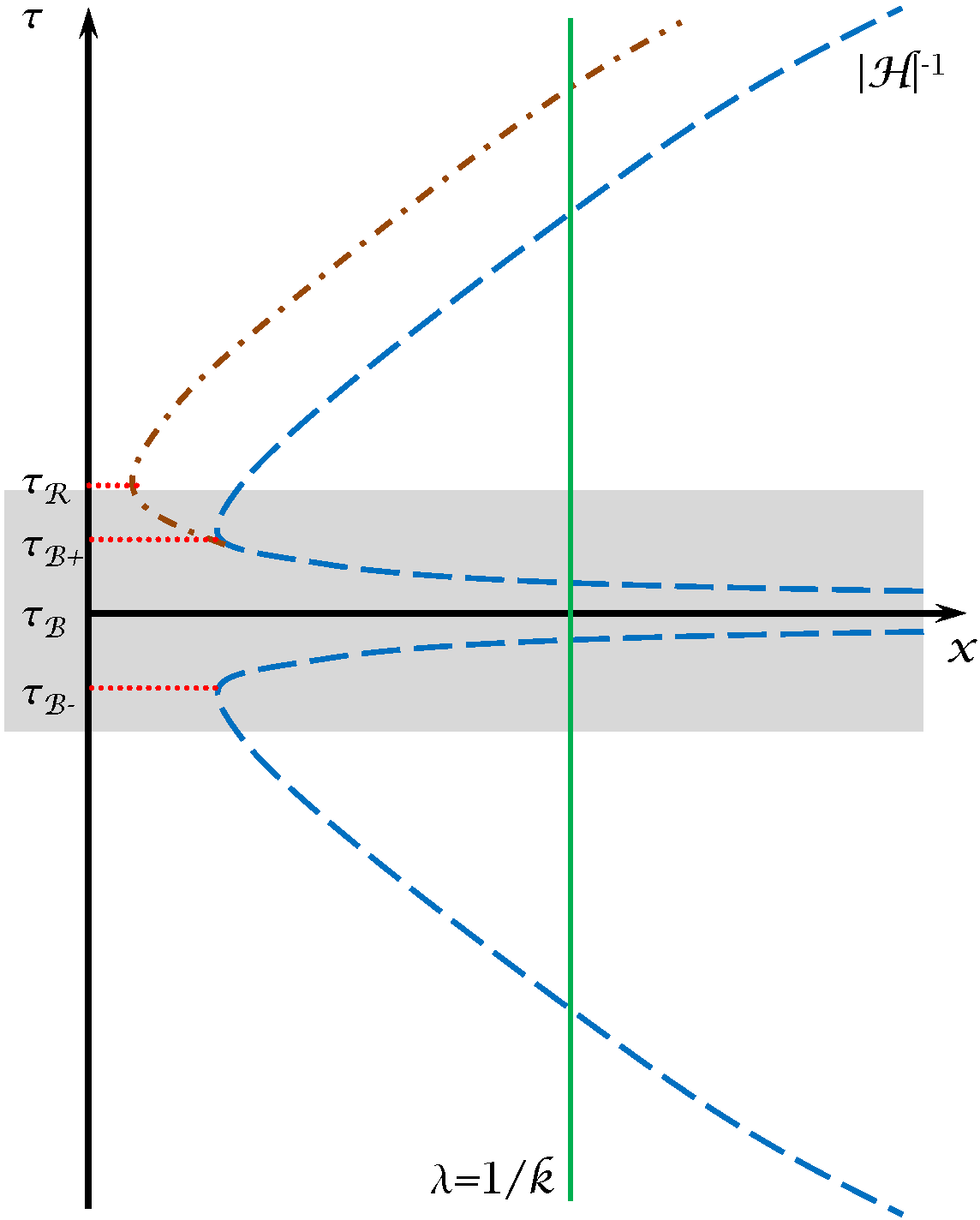}
 	\caption{ A potential depiction of a nonsingular bounce in conformal time.
 		A mode $\lambda$ first exits the Hubble horizon prior to the bounce and reenters thereafter.
 		The different times label different epochs: a nonsingular bounce phase starts at $\tau_{B-}$, ends at $\tau_{B+}$ and
 		then a reheating phase occurs until $\tau_{reh}$.
 		The blue-dashed conformal Hubble radius represents only one particular model where the background EoS is $w>0$ except during the bounce.
 		The brown dot-dashed line is another possible evolution if some energy density with EoS having $w<-1/3$ dominates the
 		evolution between the bounce and reheating. We will remain agnostic about what happens in the shaded bouncing region
 		and focus on studying the contracting phase beforehand and assume a standard radiation dominated Big Bang cosmology afterward.
 		When discussing the energy scale of the matter bounce in Sec. \ref{EscaleCont}, we will make the assumption that the uncertain shaded region, encompassing the start of the bouncing phase until the end of reheating, occurs quickly enough that it can be well approximated by a singular bounce where the scale factor is continuous but the Hubble parameter might not.}\label{bouncehistory}
 \end{figure}

Figure \ref{bouncehistory} shows the cartoon of a bouncing scenario in conformal time. One notable feature is how certain comoving modes become larger than the Hubble radius as it decreases during the contraction.
Note that the small scale modes in the CMB entered the Hubble radius during a radiation dominated period. As the matter bounce scenario requires these same modes to exit the Hubble radius during a matter dominated epoch, it implies that a perfectly symmetric bounce cannot occur. The cosmological history must therefore be asymmetric. This can be obtained by having a period of ``reheating'' in which some of the energy density is transformed into radiation \cite{Quintin:2014oea} allowing for a longer period of radiation domination after the bounce.
 \\

\subsection{Generation of perturbations in a bouncing model} \label{pertreview}
As discussed in the Introduction, fluctuations in the adiabatic direction of known bouncing models will generically yield a scalar
power spectrum of which the tilt is irreconcilable with CMB measurements. Hence one has to resort to an entropic mode which must eventually seed the fluctuations
seen today. Here we briefly review the details behind those statements. Our scenario requires two key ingredients. The first is some matter to produce the correct contraction while the second
is a homogeneous massless scalar field $\varphi$.

The scalar fluctuations can be characterized by the comoving curvature perturbation of which the Fourier mode $k$ is given by \cite{Bassett:2005xm,muktcp}
\be
\mathcal{R}_k=\psi_k-\frac{H}{\rho+p}\delta q_k ,
\ee
where $\psi_k$ is the curvature perturbation, $\rho$ is the energy density, $p$ is the pressure density and $\delta q_k$ is the scalar part of the 3-momentum.
 We can rewrite $\mathcal{R}_k$ in terms of gauge-invariant variables describing each individual fluids. Namely, defining $Q^{(i)}_k=-\delta q^{(i)}_k+\frac{\rho^{(i)}+p^{(i)}}{H}\psi_k$ we have
\be
\mathcal{R}_k=\frac{H}{\rho+p}\sum_i Q^{(i)}_k .
\ee

 For instance a massive scalar field $\chi$ with potential $V(\chi)=m^2\chi^2/2$ will have an EoS that oscillates around zero. This field can
 drive a matter contraction provided these oscillations occur on a time scale much shorter than a Hubble time. Throwing a massless scalar field $\varphi$ in the mix,
 we have a two field model which was studied in Ref. \cite{Cai:2013kja} and allows us to write
\be
\mathcal{R}_k= H\Big(\frac{ \dot{\varphi} Q^{(\varphi)}_{k}+ \dot{\chi}Q^{(\chi)}_{k}}{\dot{\varphi}^2+\dot{\chi^2}}\Big),
\ee
with $Q^{(\varphi)}_{k}=\delta \varphi_k+\frac{\dot{\varphi}}{H}\psi_k$ being the gauge-invariant Mukhanov-Sasaki variable of $\varphi$ (and $Q^{(\chi)}_{k}$ is the corresponding one for $\chi$).
Hence, as long as $\chi$ dominates the energy density $\rho\sim \dot{\chi}^2+m^2\chi^2$ we have $\dot{\chi}\gg\dot\varphi$ and the spectrum of $\mathcal{R}_k$ is set
by $Q^{(\chi)}_{k}$. This can change if $\varphi$ becomes the dominant field and reflects some of the time dependence of $\mathcal{R}_k$ if there is a significant nonadiabatic pressure component.

We define $u^{(\chi)}_{k}=aQ^{(\chi)}_{k}$ and $u^{(\varphi)}_{k}=aQ^{(\varphi)}_{k}$ as these variables
have equations of motion that can be written as \cite{muktcp},
\begin{align}
\big(u^{(\chi)}_{k}\big)''+\Big(k^2+m^2 a^2-\frac{a''}{a}\Big)u^{(\chi)}_{k}&=0,\\
\big(u^{(\varphi)}_{k}\big)''+\Big(k^2-\frac{a''}{a}\Big)u^{(\varphi)}_{k}&=0.\label{eomMassless}
\end{align}

  For the massless scalar field this is solved by the two solutions,
 \be\label{solmassless}
u^{(\varphi)}_{k}=\frac{A}{\sqrt{2k}}\text{e}^{-ik\tau}\Big(1-\frac{1}{k\tau}\Big)+\frac{B}{\sqrt{2k}}\text{e}^{ik\tau}\Big(1+\frac{1}{k\tau}\Big).
\ee
At very early times, when $k\tau\gg 1$, imposing the Minkowski vacuum state picks out the $A=1,~B=0$ solution.
At late time when $\tau k \ll 1$ this initial condition evolves to yield a scale-invariant spectrum
\be
P^{(\varphi)}_{k}\sim k^3\Big|\frac{u^{(\varphi)}_{k}}{a}\Big|^2\sim \frac{1}{\tau^6}.
\ee

The situation is more complicated for the massive field due to the $a^2 m^2$ term. If the field is to mimic matter we require that $m^2a^2\gg k^2$. Then at small scale we can impose initial conditions using the Wentzel?Kramers?Brillouin (WKB) approximation,
\be
u^{(\chi)}_{k}=\frac{1}{\sqrt{2am}}\text{e}^{-i\int^\tau a m\text{d}\tau},
\ee
which is valid if $m\gg 2/\tau$. This approximation turns out to fail when $m^2 a^2\sim a''/a$ and transition to a solution of
the form $u^{(\chi)}_{k}\sim A\tau^2 + B/\tau$ with both coefficients $A$ and $B$ set by the initial conditions which are independent of $k$.
Therefore we can conclude that the amplitudes of the modes $u^{(\chi)}_{k}$ have no $k$ dependence and the spectrum is deeply blue as it goes like $P^{(\chi)}_{k}\sim k^3$.

One can note that the initial spectrum of perturbations will depend on the details of the component responsible for the matter contraction.
To see this in more detail let us assume for simplicity that no entropy perturbation exists. The only energy component will be an unknown matter-like
fluid described by its EoS $w=p/\rho\simeq 0$ and its speed of sound $c_s^2=(\partial p/\partial\rho)|_s=\dot{p}/\dot{\rho}$.
An equation of motion can then be written for $u_k=z \mathcal{R}_k$ where $ z=a\frac{\sqrt{\rho+p}}{H c_s}$  \cite{DeFelice:2009bx},
\be
u_k''+\big(c_s^2 k^2-\frac{z''}{z}\big)u_k=0.
\ee
The properties of the fluid on the evolution of $u_k$ are captured by $z$ and the speed of sound.
This would generate a scale-invariant spectrum provided that $z\propto a$ and $\dot{c_s}/c_s \ll H$.
A scalar field with exponential potential can fulfill these properties \cite{deHaro:2015wda,Elizalde:2014uba}, but in our previous example
we had time varying EoS for $\chi$ which made $z$ not proportional to $a$.
More realistically we would expect the matter contraction to be due to some nonrelativistic perfect fluid composed of dust particles.
In this case the speed of sound grow as $c_s\propto a^{-1}$ hence $z\propto a^2$, giving a blue spectrum.

Nevertheless, even with an entropic mechanism the above models do not generate a slight red tilt.
To obtain such feature, consider a background evolving as $a\sim \tau^p$ so that $a''/a=p(p-1)/\tau^2$.
A slight deviation from perfect matter contraction, parametrized by $\epsilon\ll 1$ through $p=2(1+\epsilon)$, gives $a''/a\approx \frac{(3/2+2\epsilon)^2-1/4}{\tau^2}=\frac{v_s^2-1/4}{\tau^2}$.
This allows the spectral index of the massless scalar field to be red as it is given by $n_s-1=3-2v_s=-4\epsilon$.
The measurement of $n_s\sim 0.968\pm 0.006$ provided by the latest Planck data\cite{Ade:2015xua} determines the needed value of $\epsilon$.
A mechanism converting this entropic perturbation to an adiabatic one, such as presented in Ref. \cite{Cai:2013kja}, would provide $\mathcal{R}_k$ with the correct spectrum.

 The hope we would like to convey in this paper is that such deviation from perfect matter contraction should be due to physics that still impacts
 the cosmological evolution after the bounce and hence be tested. The logic behind the statement stems from an EFT point of view: if the matter bounce
 occurs at low energy scales-a question that will be explored in Sec. \ref{EscaleCont}-only the low energy degree of freedoms will be important to describe the dynamics.
 Such low energy DOF might only become observable again at late times, the cosmological constant being an example.
 If future observations could detect a departure from the expected $\Lambda$CDM expansion,
 it could be a nice hint for the validity of a matter bounce if the same departure during the contraction allows for a red tilt to be generated.
 This point of view naturally raises questions about the nature of the required massless isocurvature mode. For instance where is it today? In the current paper we do not attempt to give a satisfying answer to this question. It could be something along the line of a quintessence field responsible for dark energy, a
 natural candidate in the context of Sec. \ref{modelrealization} where we will consider an explicit model that realizes the red tilt.
 \\

\section{Energy Scale during Contraction}\label{EscaleCont}
In this section we argue that matter bounce models can generically have small energy density at the time the CMB modes exited the horizon.
We will make the assumption that the background evolves with an EoS $w(t)\ge 0$ that is increasing  (decreasing) toward (away from) the bounce.
This insures that the comoving Hubble radius $\mathcal{H}^{-1}$ shrinks and grows in a way similar to what is shown by the blue-dashed curve on the plot of Fig. \ref{bouncehistory}.
This would be false if the bounce itself was triggered by a null energy condition (NEC) violating fluid as it would have $w<-1$ which is smaller than $w= 0$ during matter contraction (and seen around the bounce point of of Fig. \ref{bouncehistory}).
This assumption can also break down if a phase transition-which one could expect to happen-occurred at very early time. In that case, parts of the Universe can be trapped in a false vacuum leading to topological defects. These defects can cause an inflationary stage to occur which then makes it impossible to meaningfully estimate the energy scales involved before the bounce.
As such, we want to focus on scenarios which do not have (or have an expansion close to) an inflationary period between the matter contracting phase and today.
We also have not found evidence for any type of matter (other than the cosmological constnat (CC)) with $w<0$, and it is reasonable to consider bouncing models that
do not have such ingredients dominating the background for long period of times.
This implies that our conclusions, and assumption, will be approximately correct provided the scale factor does not evolve appreciably during any times some energy component with negative EoS dominates.
For instance, we assume the bounce occurs in such a way that the scale factor is of similar size across the bounce, i.e. $a(\tau_{B-})\simeq a(\tau_{B+})$ using the notation in Fig. (\ref{bouncehistory}).

The size of the CMB's smallest length scale probed by Planck is about 100 times smaller than the baryon acoustic oscillation (BAO) scale, which is about 150 Mpc today. We denote the comoving wave number of this mode by $k_s$. It crosses the horizon twice at $k_s=a_- H_-=a_+ H_+$ when it is exiting(entering) at $-$($+$).
The largest scale in the CMB is about $10^3$ times bigger and labeled by $k_L=10^{-3} k_s$.
We parametrize our ignorance of the background between the time of the bounce and the time $k_s$ exits the Hubble radius with a constant effective EoS denoted by $w_{-}$ which is some time average of the background EoS. This allows us to relate $H_-$ to $H_B$ in $k_s=a_-H_-$ and write
\begin{align}\label{kshorizonexit}
k_s=a_-H_B\Big(\frac{a_B}{a_-}\Big)^{\frac{3}{2}(w_-+1)},
\end{align}
which implies,
\be
\frac{a_B}{a_-}=\Big(\frac{k_s}{H_B a_B}\Big)^\frac{2}{1+3w_-}.
\ee
The energy density when $k_s=a_- H_-$ is
\begin{align}
\rho^s_- \simeq M_p^2 H_-^2 &\simeq M_p^2 H_B^2 \Big(\frac{a_B}{a_-}\Big)^{3(w_-+1)} \nonumber\\
&\simeq M_p^2 H_B^2 \Big(\frac{k_s}{H_B a_B}\Big)^{6\frac{(1+w_-)}{1+3w_-}}.
\end{align}
We similarly parametrize the EoS after the bounce but before big bang nucleosynthesis (BBN) by $w_+$ in order to write
\be
H_{B+} \Big(\frac{a_B}{a_{BBN}}\Big)^{3(1+w_{+})/2}=H{_{BBN}} \,.
\ee
This allows us to express $a_B$ in terms of the Hubble scale right after the bounce $H_{B+}$ and at BBN $H_{BBN}$.

Moreover the energy scale after reheating must be higher than $\rho_{BBN}\simeq \text{MeV}^4$ but must be below the energy scale of the bounce, and therefore $H_B\ge H_{B+}\ge H_{BBN}$.
Assuming $w_+\ge0$ implies $\rho^s_-$ is maximized by having $H_{B+}$ as large as possible, we set $H_{B+}=H_B$ to get an upper bound. With this we find

\begin{align}
\rho^s_- &\lesssim M_p^2 H_B^{\frac{4(w_{-}-w_+)}{(1+3w_{-})(1+w_+)}}H_{BBN}^{\frac{-4(1+w_-)}{(1+3w_{-})(1+w_+)}} \Big(\frac{k_s}{ a_{BBN}}\Big)^{6\frac{(1+w_-)}{1+3w_-}}\nonumber\\
&\simeq \Big(\frac{M_p k_s}{ a_{BBN}}\Big)^{6\frac{(1+w_-)}{1+3w_-}} \big(E_{BBN}^{-1-w_-}E_B^{w_--w_+}\big)^{\frac{8}{(1+w_+)(1+3w_-)}} ,\label{rhobound}
\end{align}
where in the second line we have rewritten $H_B$ and $H_{BBN}$ in terms of the energy scale using $M_p^2 H^2\simeq E^4$.
We can now plug in some numbers to see what we could expect from different types of bounce models.
With $a_0$ being the scale factor today, we have the following known values for the parameters in Eq. (\ref{rhobound}),
\begin{align}\label{parameters}
 & k_s/a_0 \sim \text{Mpc} \sim 10^{-38} \text{GeV} ,~~~ a_{BBN}/a_0\sim 10^{-8},\nonumber\\
 & E_{BBN} \sim 10^{-3} \text{GeV} ,~~~ M_p \sim 2.4\times 10^{18}\text{GeV}.
\end{align}

\begin{figure}[tbhp]
	\centering
	\includegraphics[width=0.45\textwidth]{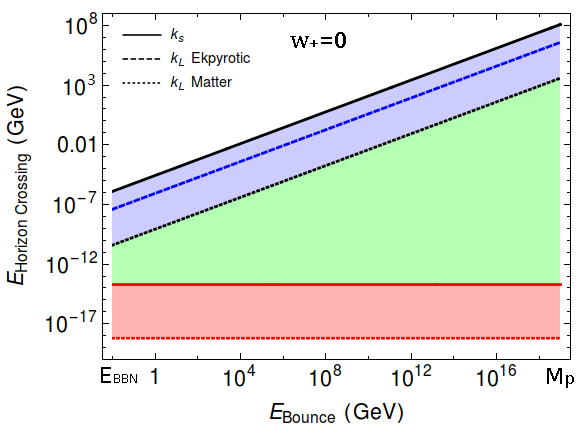}\\
	\includegraphics[width=0.45\textwidth]{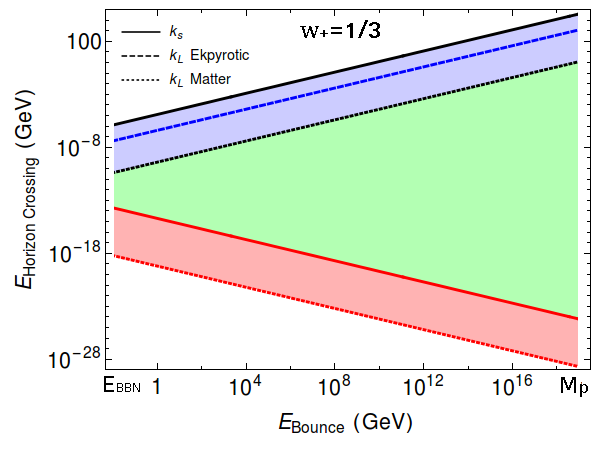}
	\caption{A plot of an upper bound on the energy scale during Hubble crossing of the CMB modes in the contracting phase as a function of the energy scale at the bounce.
		The solid lines represent the smallest mode $k_s$ while the dotted ones represent the largest mode $k_L$. The shaded region characterizes our ignorance on the effective background EoS $w_-$ from the
		time $k_s$ crosses the horizon until the bounce. The top solid line occurs when $w_-\gg 1$ while the bottom solid line is when $w_-=0$. We expect any matter bounce models to fall
		in the green region that interpolates between these two extremes. These plots do depend on the effective EoS after the bounce and BBN which we denote by $w_+$.
		See the text for additional details.}\label{edensfig}
\end{figure}

Three numbers remain unknown: $w_-$, $w_+$ and $E_B$. As $k_s$ reenters the Hubble radius during the radiation epoch we
might expect $w_+$ to be close to $1/3$. We will then consider two cases to represent Eq.(\ref{rhobound}) : one with $w_+=0$ and the other with $w_+=1/3$. Note that the upper bound becomes smaller as $w_+$ increases.
As a function of the energy scale of the bounce $E_B$, Fig. \ref{edensfig} shows the upper bound on the energy scale during first horizon crossing $E_k=(\rho_-(k))^{1/4}$
for various modes $k$ and effective EoS $w_-$.
The solid black line represents the energy scale at the time $k_s$ exits the horizon if $w_-\gg 1$,
i.e. an Ekpyrotic phase takes place between the bounce and $\tau_-$ when $k_s=a_- H_-$.
If the CMB scales leave the horizon solely during an Ekpyrotic phase (i.e., we do not have a matter bounce scenario) we still can compute $E_{k_L}$ which is shown by the dashed blue line.
We see that large numbers such as $E_{k_s}\sim 10^5 \text{GeV}$ can be achieved if $\rho_B$ occurs near the Planck scale.
On the other hand a matter bounce model would require a matter dominated contraction prior to $\tau_-$. This would yield a $E_{k_L}$ given by the dotted black line which is at most on the order of a $\text{GeV}$.
Therefore, in models with a very long Ekpyrotic phase we expect $E_k$ to be quite large and independent of $w_+$.
Of course it is sensible to expect the matter dominated phase to last for some time after $\tau_-$, and hence at the other extreme we can have models with a matter domination epoch
for most of the contracting phase with $w_-\simeq 0$. This yields a $E_{k_s}$($E_{k_L}$) shown by the solid(dotted) red line. The energy scales are much lower and take values as low as $10^{-28}$ GeV for $w_+=1/3$ and $10^{-18}$ GeV
for $w_+=0$. Interpolating between the two extremes of $w_-$ we will find that any specific model should lie in the green region. Nevertheless, this identifies the existence of a large parameter space which has matter bounce models
which generate the statistical properties of the CMB fluctuations at very low energies. These numbers should be compared to the current
energy density today, $\rho_0\simeq (10^{-12} \text{GeV})^4$.

What happens if $w_{+}$ is negative? The power of $E_B$ is given by $\frac{8(w_--w_+)}{(1+w_+)(1+3w_-)}$ and so lowering $w_{+}$ will increase the bound on $\rho^s_-$ to render it meaningless.
Models that have such features push $\rho^s_-$ to large values. However, as no topological defects have been detected, if they were produced they must have decayed into radiation at very early times.
Additionally, the monopole problem could impose bounds on the bounce scale which could further reduce the parameter space of Fig. \ref{edensfig}.
Hence we restate that such components can exist and yet still impose a strong lower upper bound on $\rho^s_-$ as long as the defects do not dominate the evolution for a significant amount of time.
It would be interesting to see how these conclusions are affected when analyzing models with such features.

\section{ CMB Red tilt}\label{sec_redtilt}

We have mentioned earlier that a slight deviation from matter contraction can produce a small tilt in the power spectrum of a massless scalar field.
A red tilt is generated if an energy density with $w<0$ dominates during the contracting phase. To produce a slight red tilt, one might consider pressureless matter $w=0$ to be the dominant contribution and some subdominant energy component with $w<0$ so that the effective $w$ is slightly less than, but still very close to, zero.
 Some possibilites of energy densities having this behavior includes: the CC with $w=-1$, defects such as domain walls with $w=-2/3$ and cosmic strings with $w=-1/3$. The last two examples might create additional difficulties for the
homogeneity of the background if they become significant when approaching the bounce, for instance a network of intersecting strings would eventually evolve as an inhomogeneous radiation fluid \cite{Avelino:2002xy}.
It has been argued that a cosmological constant could give rise to the desired tilt \cite{Cai:2014jla}; however, we demonstrated in Sec. \ref{LCDMscen} that the desired
tilt cannot be maintained for a wide enough range of $k$ as there will be significant running. This issue can be avoided by
using an interacting dark energy model which we explore in Sec. \ref{modelrealization}.

\subsection{Matter contraction with noninteracting dark energy (the $\Lambda$CDM bounce scenario)}\label{LCDMscen}
Assuming the EoS of dark energy is $p_d=w\rho_d$,  the Friedmann equation is
\be \label{friedmann}
H^2=\frac{1}{3M_{p}^2} \left( \rho_m+\rho_d \right)~,
\ee
with $\rho_m=\rho_{m0}a^{-3}$ and $\rho_d=\rho_{d0}a^{-3(1+w)}$.
For $w=-1$  we return to the $\Lambda$CDM bounce scenario discussed in Ref. \cite{Cai:2014jla}.
We introduce the ratio of DM and DE energy density $\varrho\equiv\rho_m/\rho_d$ and define $\varrho_0\equiv\rho_{m0}/\rho_{d0}$, so that $\varrho=\varrho_0a^{3w}$.
In conformal time, Eq.(\ref{friedmann}) can be expressed as
\be \label{fr1}
a'(\tau)=\frac{1}{\sqrt{3}M_{p}}\sqrt{a\rho_{m0}\left(1+\frac{1}{\varrho}\right)}~.
\ee
Taking a derivative with respect to $\tau$,
\be
a''(\tau)=\frac{1}{6M_p^2}\rho_{m0}\left(1+\frac{1-3w}{\varrho}\right)~.
\ee
As $\varrho\gg 1$ in the matter dominated stage, we can solve for $a$ to first order in $1/\varrho$,
\be
a\simeq \frac{\rho_{m0}}{12M_p^2}\tau^2(1+(1-3w)/\varrho) \,,
\ee
which leads to the following expression for the ratio $a''/a$,
\be
\frac{a''}{a}\simeq\frac{2}{\tau^2}\Big(1+\frac{2(1-3w)^2-1}{2(1-3w)\varrho}+\mathcal{O}(1/\varrho^2)\Big)\simeq\frac{v_s^2-\frac{1}{4}}{\tau^2}~,
\ee
with $v_s\simeq\frac{3}{2}+\frac{2(1-3w)^2-1}{3(1-3w)\varrho}$.
Therefore, as shown in Sec. \ref{pertreview}, the spectral index of our massless field is estimated to be,
\be
n_s-1=3-2v_s=-\frac{4(1-3w)^2-1}{3(1-3w)\varrho}.
\ee
For $w<0$, this gives the desired red tilt.
Note that because $\varrho=\varrho_0 a_0^{3w}\tau^{6w}$ is time dependent, the spectral index should be calculated at the time of horizon exit $|\tau_k|\sim 1/k$ of each mode. Thus we find
\be
n_s-1\sim -\frac{4(1-3w)^2-1}{3(1-3w)\varrho_0a_0^{3w}}k^{6w},
\ee
which is heavily dependent on $k$ for $w\simeq -1$. In this case, the spectral running is
\be
\alpha_s=\frac{dn_s}{d\ln k}=-6 (n_s-1),
\ee
and is larger than the absolute value of $n_s-1$ by a factor of 6. The current observational bounds on $\alpha_s$ are smaller than $n_s-1$ by about 1 order of magnitude \cite{Ade:2015xua}; therefore, the $\Lambda$CDM bounce scenario cannot generate the observed cosmological perturbations.

\subsection{Matter contraction with interacting dark energy}\label{modelrealization}

We now present a mechanism to generate a red power spectrum with little running. The mechanism relies on the introduction of a  dark matter and dark energy interaction.
Such interactions in the dark sector have been considered previously in the literature \cite{chimento,intde1,Comelli:2003cv,Zhang:2005rg,Cai:2004dk,Guo:2004xx} as an attempt to explain the coincidence problem and are precisely of the form needed to generate a slight red tilt with little running in a contracting universe.

\begin{figure}[tbhp]
\centering
\includegraphics[width=0.9\linewidth]{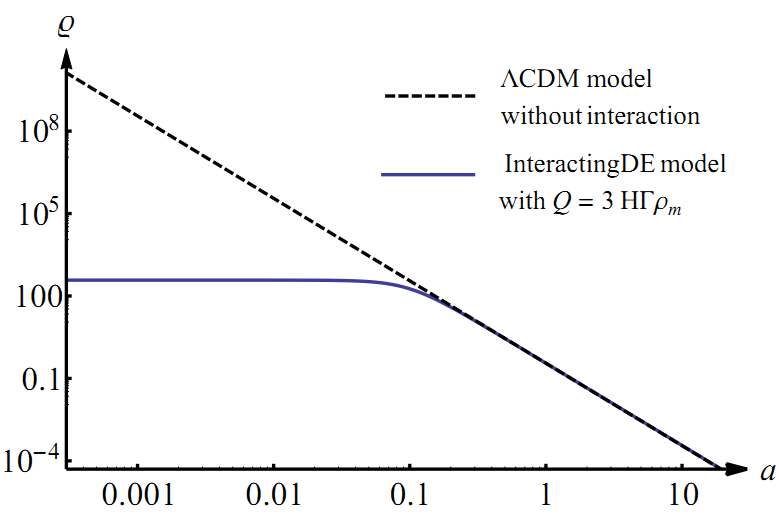}
\caption{The evolution of the DM to DE energy density ratio $\varrho\equiv\rho_m/\rho_d$ in two different models. Here we set $\Gamma=0.0026$ and $w=-1$.}
\label{rafig}
\end{figure}

Consider a phenomenological model with dark energy and matter being two fluids having energy-momentum tensor $T^{\mu\nu}_d$ and $T^{\mu\nu}_m$ respectively. By Einstein equations and the Bianchi identity, the total energy-momentum tensor is conserved,
\be
0=\nabla_\mu T^{\mu\nu}=\nabla_\mu  T^{\mu\nu}_d+\nabla_\mu  T^{\mu\nu}_m= Q^\nu_d+Q^\nu_m~.
\ee
Here, $\nabla_\mu T^{\mu\nu}_i=Q^\nu_i$ is non-zero whenever interactions are present. Thus the energy transfer satisfies $Q_m^0=-Q_d^0\equiv Q$.
While the background evolution remains adequately described by Eq.(\ref{friedmann}). The energy conservation equations for matter and dark energy are expressed as
\begin{align}
\dot{\rho}_m+3 H \rho_m=Q~, \label{EOMCDM}\\
\dot{\rho}_d+3 H (\rho_d+p_d)=-Q~.\label{EOMDE}
\end{align}
Here we have used $p_m=0$ for the matter component and assume a constant EoS $w=p_d/\rho_d\simeq -1$ for dark energy.~\footnote{We
will have to chose the dark energy EoS to be near, but slightly less than, $-1$ to insure stability \cite{He:2008si}.}
Thus for $Q>0$, energy flows from dark energy to matter, and for $Q<0$ energy flows from matter to dark energy.
Using Eqs.(\ref{EOMCDM}) and (\ref{EOMDE}), the ratio of matter and dark energy density $\varrho \equiv\rho_m/\rho_d$ evolves as
\be \label{rt}
\frac{d\varrho}{dt}=3H\varrho\left[w+\frac{Q}{9H^3M_{p}^2}\frac{(1+\varrho)^2}{\varrho}\right]~.
\ee
Unlike the $\Lambda$CDM universe corresponding to $Q=0$, this expression allows for a nontrivial fixed point if $Q\propto H^3$.
In what follows we consider one of the simplest models of interacting dark energy that have this feature, namely $Q=3H\Gamma\rho_m$ with a constant $\Gamma>0$.  In this case, Eqs. (\ref{EOMCDM}) and (\ref{EOMDE}) become
\be \label{dm}
\frac{d\rho_m}{da}+3(1-\Gamma)\frac{\rho_m}{a}=0~,
\ee
\be \label{de}
\frac{d\rho_d}{da}+3(1+w)\frac{\rho_d}{a}+3\Gamma\frac{\rho_m}{a}=0~.
\ee
Here we used the scale factor $a$ instead of the cosmic time $t$ as it is more convenient to describe the evolution before the bounce.
Using this new time coordinate, Eq.(\ref{rt}) becomes
\be \label{ra}
\frac{d\varrho}{da}=\frac{3}{a}\left[\Gamma \varrho^2+(\Gamma+w)\varrho\right]~.
\ee
This equation can be solved analytically. Setting the initial condition $\varrho=\varrho_0$ at $a=1$, we have
\be \label{ras}
\varrho(a)=\frac{\varrho_0(w+\Gamma)}{(w+\Gamma+\varrho_0\Gamma)a^{-3(w+\Gamma)}-\varrho_0\Gamma}~.
\ee
In Fig.\ref{rafig} we compare $\varrho(a)$ for an interacting dark energy model against the $\Lambda$CDM prediction and show how the fixed point is approached at $\varrho=\varrho^+$ for $a\ll 1$.
The value of $\varrho^+$ is obtained using Eq.(\ref{ra}) and the condition $\frac{d\varrho}{da}=0$. We find two constant solutions of $\varrho$,
\be
\varrho^+=-\frac{\Gamma+w}{\Gamma}~~~\text{and}~~~\varrho^-=0~.
\ee
Thus in a situation with $\Gamma\ll1$, we have $\varrho\simeq\rho+\gg1$ for $a\ll1$ and this nearly constant stage is matter dominated.

In what follows, we show how a slight red tilt for $P_\varphi(k)$ is produced from this interacting dark energy model.
Solving Eq.(\ref{dm}) we find $\rho_m\propto a^{-3(1-\Gamma)}$ and using this with the first Friedmann equation (\ref{friedmann}) in the regime where $\varrho$ is
approximately constant, we find $a(t)\propto t^{2/3(1-\Gamma)}\propto \tau^{2/(1-3\Gamma)}$. The perturbations of $\varphi$ are again be given by Eq. (\ref{eomMassless}) with
\be
\frac{a''}{a}
=\frac{\left(\frac{2}{1-3\Gamma}-\frac{1}{2}\right)^2-1/4}{\tau^2}=\frac{v_s^2-1/4}{\tau^2}.
\ee
The spectral index is therefore, $n_s-1=3-2v_s=4-\frac{4}{1-3\Gamma}$.
The latest Planck measurements \cite{Ade:2015xua} set $n_s=0.968\pm0.006$ which fixes $\Gamma=0.0026\pm0.0005$ in order to get the measured red tilt. Another feature, arising due to being near the fix point, is that the running of $n_s$ will be small and negative, and the exact value is related to the time variation of $\varrho$.\\

Having such an energy transfer will also affect the late-time Universe by altering the cosmological expansion.
Therefore this model can be, and has been, tested by late-time observations \cite{Costa:2013sva,Abdalla:2014cla}.
The analysis of Ref. \cite{Costa:2013sva} fits $\Gamma=0.002272^{+0.00103}_{-0.00137}$  at the 68\% confidence level for Planck+BAO data,
and $\Gamma=0.001494^{+0.00065}_{-0.00116}$  at the 68\% confidence level for Planck+BAO+SNIa+H0 data.
There are still large uncertainties in these constraints, but the fit suggests a 2 sigma deviation from $\Lambda$CDM and an agreement with the demands to generate the proper red tilt.
Improvements on measurements in the not so distant future will further determine the viability of such scenarios.

\section{Conclusion}\label{conclusion}
As an alternative to the simple picture given by the inflationary scenario,  current matter bounce models require many intricacies to agree with observations.
The original idea of having a single massless scalar field responsible for the CMB perturbations cannot explain the observed red tilt. Additional parameters must be added to achieve these features. This presents a difficult challenge for the matter bounce scenario,
however, if the new parameters are expected to exist, observational opportunities are also present. This is where a particularly interesting feature of the matter bounce comes into play:
the energy density of the Universe at the time the statistical properties of the perturbations are frozen in can be very low, in some cases many orders of magnitude smaller than today's energy density.
 Hence, unlike other pre-Big Bang or early Universe scenarios, only low energy degrees of freedom are relevant during the times of CMB mode horizon crossing. The observations that uncover those degrees of freedom are made at low redshifts where their effects are most easily seen. The cosmological constant is an example: it is a parameter of the theory that only becomes
important during the late stages of our Universe. In a universe where the matter bounce occurred, this opens the possibility
of determining the identity of the extra parameters that  make up a viable model, especially when considering the recent cosmological hints of modified gravity \cite{DiValentino:2015bja}.
To illustrate how the hunt for these parameters might ensue, we note that a
deviation from perfect matter contraction would give a red tilt to perturbations. Whatever is responsible for this deviation
is likely to also cause a similar effect after the bounce and can be observationally detected. The introduction of a coupling in the dark sector
that allows energy to flow from dark matter to dark energy is able to support a contracting phase giving the correct $n_s$.
These interacting models have been considered as extensions of $\Lambda$CDM \cite{Costa:2013sva,Abdalla:2014cla} and the best fit to the Planck+BAO+SNIa+H0 data has the coupling $\Gamma$ deviate from zero at the 2 sigma confidence level. Intriguingly, the strength of the coupling needed to set the measured CMB value of $n_s$ sits comfortably within the error bars found by this analysis.

It is interesting to speculate further about the potential new signatures that might arise in viable matter bounce models. For instance,
another issue faced by simple matter bounce scenarios is to realize a small tensor-to-scalar ratio consistent with CMB observations, and
this problem has been comprehensively reviewed in Ref. \cite{Cai:2014bea}.
However, if future low redshift observations could
detect a small graviton mass \cite{deRham:2014zqa}, this feature would change the details of the tensor modes vacuum state and could affect the predictions for the tensor-to-scalar ratio.
The details are left for future work, but the role of a massive graviton in matter bounce models is yet another example of a low energy DOF that can be searched for.

Additionally, probing the matter bounce using low redshift observations could also shed light on experimental tests of nonsingular bounces by other mechanisms, such as by using the direct dark matter searches or the BBN bound as analyzed in Refs.\cite{Cheung:2014pea,Li:2014era,Li:2015egy,Cheung:2014nxi}.

\section{Acknowledgements}
We are grateful to Elisa Ferreira, Jerome Quintin, Jeff Hyde and Yao Ji for useful discussions and comments. Y.F.C. and D.G.W. are supported in part by the Chinese National Youth Thousand Talents Program and by the USTC startup funding (Grant No.~KY2030000049) and by the National Natural Science Foundation of China (Grant No. 11421303). F.D. also thanks USTC for the hospitality and funds during the initial stages of this project. F.D. is supported by a B2 scholarship from le Fonds de Recherche du Qu\'ebec\textendash Nature et Technologies (FRQNT).

\end{document}